\documentclass[11pt,a4paper,titlepage,aps,superscriptaddress,preprint,pra]{revtex4-1}
\usepackage{hyperref}
\usepackage{amsmath}
\usepackage{color}
\usepackage{graphicx}

\begin{document}

\title{Counter-propagating dual--trap optical tweezers based on linear momentum conservation}

\author{M. Ribezzi--Crivellari}
\affiliation{Departament de Fisica Fonamental, Universitat de Barcelona}
\author{J. M. Huguet}
\affiliation{Departament de Fisica Fonamental, Universitat de Barcelona}	  
\author{F. Ritort}
\affiliation{Departament de Fisica Fonamental, Universitat de Barcelona}
\affiliation{Ciber-BBN de Bioingeneria, Biomateriales y Nanomedicina}

\date{today}


\begin{abstract}
We present a dual-trap optical tweezers setup which directly measures forces using linear momentum conservation.
The setup uses a counter-propagating geometry, which allows momentum measurement on each beam separately. 
The experimental advantages of this setup include low drift due to all-optical manipulation, and a robust calibration 
(independent of the features of the trapped object or buffer medium) due to the force measurement method.
Although this design does not attain the high-resolution of some co-propagating setups, we show that it can be used to perform
different single molecule measurements: fluctuation-based molecular stiffness characterization at different forces and hopping experiments on molecular hairpins.
Remarkably, in our setup it is possible to manipulate very short tethers (such as molecular hairpins with short handles) down to the limit where beads are almost
in contact.
The setup is used to illustrate a novel method for measuring the stiffness of optical traps and tethers on the basis of equilibrium force fluctuations,
i.e. without the need of measuring the force vs molecular extension curve. This method is of general interest for dual trap optical tweezers
setups and can be extended to setups which do not directly measure forces. 
\end{abstract}
\maketitle


\section{Introduction}

\begin{center}
\begin{figure}
\centering
\includegraphics[width=0.4\textwidth]{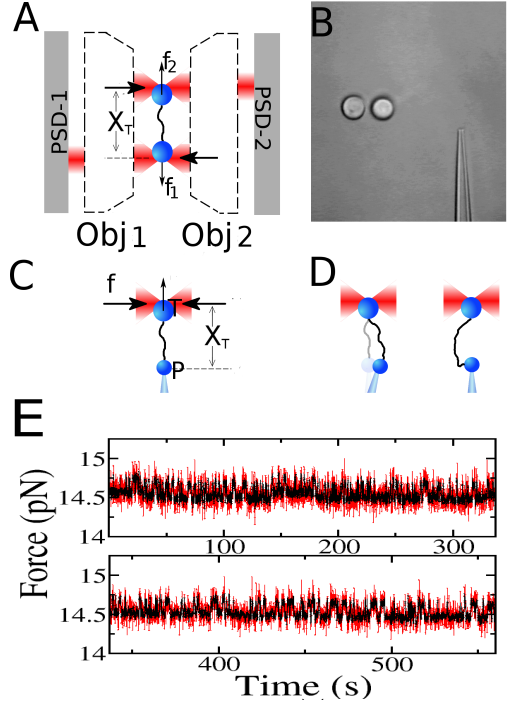}
\caption{\label{Fig:1} {\bf DT vs ST setup.} A) The experimental DT setup described in this paper, based on the MiniTweezers \cite{PNAS.huguet.2010}. 
Two optical 
traps are created with two objectives (Obj 1, Obj 2) from counter-propagating laser beams. The force
exerted by the traps is measured by two Position Sensitive Devices (PSD 1, PSD 2) with sub-picoNewton resolution by direct measurement of light momentum, while the position of both
traps is monitored with nanometer accuracy. The horizontal black arrows show the direction of light propagation. Two force signals, $f_1,f_2$ are measured (vertical black arrows). 
B) Two optically trapped beads and the micropipette used in the ST setup in a video-microscopy image. The beads are $\simeq$ 4 $\mu$m in diameter.
C) The ST setup introduced in \cite{MEnzy.smith.2003}
is used to create a single--trap and to manipulate a molecule tethered between two beads. One bead is optically trapped, while the other is immobilized on the tip
of a micropipette by air suction.
D) Experimental issues in the ST setup. Left panel: drift effects due to uncontrolled movements of the micro--pipette. Right panel: the bead in the pipette is not free to rotate and
the molecule can be misaligned with respect to the pulling direction. These effects are largely reduced in the DT setup. 
E) The great stability of the two-trap setup allows drift free long--term measurements. Here we show a ten-minute passive mode hopping trace for a DNA hairpin
as in Section \ref{mfor} (red points: raw data acquired at 1 kHz, black points box average to 10Hz), in which the relative position of the traps drifts less than 2 nm.}
\end{figure}
\end{center}

Optical tweezers (OT) have become a key tool in the fields of biological and statistical physics \cite{Biochem.moffitt.2008,RevSciInst.neuman.2004}. OT use focused laser beams to 
form optical traps around dielectric objects.
Over the past years several important achievements in the field of single-molecule biophysics have been obtained using Optical Tweezers in the Dual Trap (DT) setup, which manipulate single
molecules tethered between two optically trapped dielectric beads (see Fig. \ref{Fig:1} A). 
DT setups have two important advantages.  On the one hand the possibility of manipulating single molecules by all-optical  means guarantees an exceptional isolation from ambient noise. 
On the other hand, in many experimental situations, the measurement of cross correlations between signals coming from two traps allows
to overcome the resolution limit imposed by the stiffness of the traps \cite{PNAS.moffitt.2006}.
Most DT setups use single beam trapping. In this case a dielectric bead is trapped near to the focus of a convergent laser 
beam by balancing the scattering and gradient forces, exerted along the optical axis.
Several experiments have been performed with this setup, e.g. the direct measurement of hydrodynamic correlations 
between trapped particles \cite{Prl.meiners.1999,JChemPhys.crocker.1997,PhilTransRoySoc.bartlett.2001}, of the stiffness of long double-stranded DNA molecules 
\cite{Prl.meiners.2000} 
and of the sequence-dependent free energy landscape of DNA hairpins and proteins \cite{PNAS.woodside.2006,Science.woodside.2006,PNAS.gebhardt.2010}.
In most cases the force applied by the trap on the sample is measured by modeling the trap as an harmonic spring. 
The force applied by the trap, $f$, is given by $f=-k x$  where $k$ is the stiffness of the equivalent spring and $x$ is 
the position of the trapped bead with respect to the center of the trap.
The stiffness of such spring can be measured in several ways by applying a known force to the trapped bead and measuring its displacement from the 
center of the trap. Unfortunately the stiffness, and thus the calibration of force measurement, depends on the details of the experimental setup, such as the size or shape 
of the trapped bead, the index of refraction of bead and surrounding medium and laser power. Moreover the harmonic model of the trap can be largely inaccurate and nonlinear
effects can appear even at moderate forces, leading to large errors in force calibration.
An exception to this rule is given by setups which measure forces based on the conservation of linear momentum \cite{MEnzy.smith.2003}. In these cases there 
is no need for trap modeling and calibration is more robust. Direct force measurement methods have already been implemented in single--trap (ST) optical tweezers setups and used in single molecule experiments
\cite{Science.smith.1996}. 
In this paper we report the performance of a novel DT setup based on single beam trapping which exploits the advantages of direct force measurements together with those 
of all-optical manipulation. The outline of this setup is shown in Fig. \ref{Fig:1} A, while in Fig. \ref{Fig:1} B we show a video microscopy image of two optically trapped
beads. This setup has been implemented in a miniaturized version of ST setup that uses counter-propagating beams
\cite{PNAS.huguet.2010} (Fig. \ref{Fig:1} C).

Single beam trapping requires large Numerical Aperture (NA) beams, to enhance the gradient force. On the contrary, direct force measurement requires low NA beams.
A strategy to implement direct force measurements in single beam traps is to increase the angular acceptance of the collector objective, which requires a specific optical design and great care in performing experiments.
This approach has been pursued in \cite{Optexp.farre.2010}. Our approach is different. The trapping efficiency is known to strongly depend both
on the NA of the beam and on the effective refraction index, i.e., the ratio of the refraction index
of the trapped bead to that of the surrounding medium. The effect of this latter variable has not been tested in \cite{MEnzy.smith.2003}.
We use underfilling (low NA $\simeq 0.6$) beams in high NA ($\simeq 1.2$) objectives.  Using underfilling beams the trap performance at effective refractive index 1.2 (polystyrene in water) is too poor to perform experiments. However by using beads with lower effective refraction index ($\simeq 1.1$) it is possible to perform single molecule experiments. Such conditions can be achieved with
silica beads in water or polystyrene beads in sucrose solution.
Even in these conditions the trap stiffness per Watt is low ($\simeq 0.4$ pN/(nm W)) compared to other setups, but sufficient to perform single-molecule experiments on small hairpins.
Contrary to other dual-trap setups, which use polarization optics to distinguish the light from the two beams we employ a counter-propagating geometry.
In this geometry the two beams 
forming the traps leave the sample region in opposite directions (horizontal black arrows in Fig. \ref{Fig:1} A) so that the light coming from each trap can be separately 
detected and the force exerted by each trap measured. In this way we avoid systematic errors coming from cross talk \cite{RevSciInst.mangeol.2008} and depolarization, but incur in reflection effects as detailed in section \ref{fff}. 
This setup works at low laser power (50 mW in the sample region), which limits trap stiffness but leads to several experimental advantages such as 
reduced heating in the sample region, reduced tether photodamage \cite{BiophysJ.landry.2009} and negligible optical binding effects between the two trapped beads \cite{Prl.burns.1989}. In addition, with our DT setup it is possible to perform single molecule experiments on very short tethers (as short as 20 nm). 
To the best of our knowledge this has not been reported in co-propagating setups.
The DT setup has some advantages over the ST setup, mainly because the latter employs a micropipette which drifts during experiments (Fig. 1D).  Drift is greatly reduced in the DT setup, although it lacks the compensation of drift between 
the two traps so characteristic of co-propagating setups. In Fig \ref{Fig:1} E a force trace from a hopping experiment (see Section \ref{mfor}) of almost 10 minutes of duration is shown. 
Moreover the ST setup only measures the position of the trap, assuming the bead in the pipette is fixed 
(Fig. \ref{Fig:1} C),
while the DT setup can measure the position of both traps, their relative distance being the control parameter in the experiments (Fig. \ref{Fig:1} A). 
The main limits of the design described in this article are the limited trap stiffness due to the strict requirements for direct force measurement and 
the misalignment of the traps along the optical axis.
As far as misalignment is concerned, we show that, although it can have a large effect on high-bandwidth elastic fluctuation measurements \cite{BiophysJ.ribezzi.2012},
it can be generally disregarded when working at lower bandwidth as in hopping or pulling experiments.

\section{Experimental setup}\label{set}

\begin{figure}
 \centering
\includegraphics[width=0.7\textwidth]{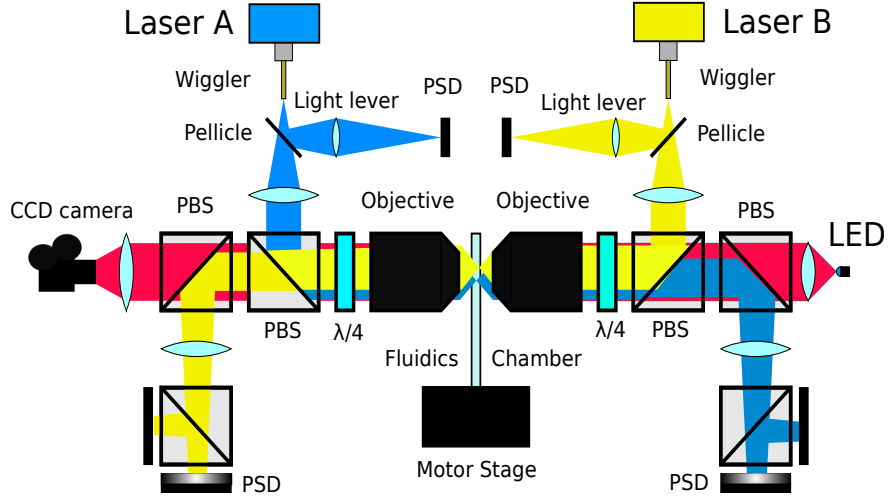}
\caption{\label{Figure:opts}
{\bf Experimental Setup.} The scheme of the optical setup, with the optical paths of the lasers (blue and yellow) and of the led (red). Fiber-coupled diode
lasers are focused inside a fluidics chamber to form optical traps using underfilling beams in high NA objectives.  All the light leaving from the trap is
collected by a second objective and sent to a Position Sensitive Detector which integrates the light momentum flux, measuring changes in light momentum \cite{MEnzy.smith.2003}.
The laser beams share part of their optical paths and are separated by polarization. Part of the laser light ($\simeq$ 5\%) is deviated by a pellicle 
before focusing and used to monitor the trap position (Light Lever). The trap is moved by pushing the tip of the fiber tip by piezo actuators (wiggler).}
\end{figure}

The optical tweezers instrument shown in Figure \ref{Figure:opts} is based on that designed by Smith et al. \cite{MEnzy.smith.2003} 
and described in \cite{PNAS.huguet.2010}.
Two optical traps are created from two different 845 nm, single mode, fiber coupled laser diodes (Lumix LU0845M200). The laser power is 130 mW, while the power of
the laser actually reaching the trap is 50 mW in all the experiments. 
These sources emit linearly polarized beams in the $\text{TEM}_{00}$ mode. 
Part of the light intensity emitted from a source (Laser A) is used to actually create the trap
while a small fraction (~5\%) is redirected by a beam splitter pellicle and condensed to a Position Sensitive Device (PSD) "Light Lever" (Fig.\ref{Figure:opts}) in order to monitor trap displacements.
Piezoelectric crystals are used to gently push the tip of the optical fiber coupled to the laser source, allowing to redirect the laser beam and move the optical trap in the optical plane (i.e. perpendicular to the optical axis). 
The light used to create the trap is set to circular polarization. This is obtained using a Polarizing Beam Splitter (PBS), which selects the horizontally polarized light, 
and a $\lambda/4$ plate. 
The beam is then focused to form the trap by a water immersion objective (Olympus UPLSAPO 60$\times$W). The light leaving the trap is then recollected by a
second objective, symmetrically positioned with respect to the first one, converted to vertical linear polarization through a second $\lambda/4$ plate and redirected by a PBS to a PSD which 
measures the momentum flux carried by the light coming out of the trap. The light from the second source (Laser Diode B) undergoes a mirror symmetric path.
The two beams share their optical path between the objectives.
In order to correctly measure forces by linear momentum conservation it is necessary to recollect all the light
exiting from the trap, which can be achieved by using underfilling beams in large numerical aperture objectives, so that the only light loss is due to reflection by mirrors and beads which
is below 2\% in the range of forces explored by this instrument (see below).
The counter-propagating geometry ensures that the light forming the traps leave the sample region in opposite directions. The light coming from each trap can be separately detected and the force 
exerted by each trap measured.
A blue LED and a CCD camera (Watec WAT-902H3 SUPREME EIA) are used to monitor the experiment by video microscopy: a lens projects the image of the focal plane to the CCD 
camera which responds both in the visible and in the infrared so that an image of both the beams and the trapped beads can be distinguished.
The instrument is controlled by a personal computer which is also used for data acquisition at 4 kHz. 

\section{Force Measurement Calibration}\label{cali}
Force measurement is obtained by collecting all the light deflected from the trapped bead \cite{MEnzy.smith.2003}.
The instrument measures forces acting in the optical plane ($x-y$ plane) upon the trapped bead using linear momentum conservation. Forces along the optical axis ($z$ direction), 
despite being measured in the single--trap setup described in \cite{MEnzy.smith.2003}, are not measured in the DT setup. 
The change in the $x,y$ components of the outgoing momentum flux (Fig \ref{Fig:califorcmet}A) is detected by using Position Sensitive Detectors (PSDs), which emit two current signals, $I_x,I_y$ proportional to the forces along the 
two orthogonal directions (Fig. \ref{Fig:califorcmet} B). Throughout the paper the tether is taken to be aligned along the $y$ axis which is the direction defined by the straight line joining the centers of the two traps.
We will thus focus our attention on the $y$ component of the force.
Calibration of the instrument amounts to the measurement of the conversion factor, $\lambda_i,\,i=x,y$, between force and current:
\begin{figure}  
\centering
\includegraphics[width=0.4\textwidth]{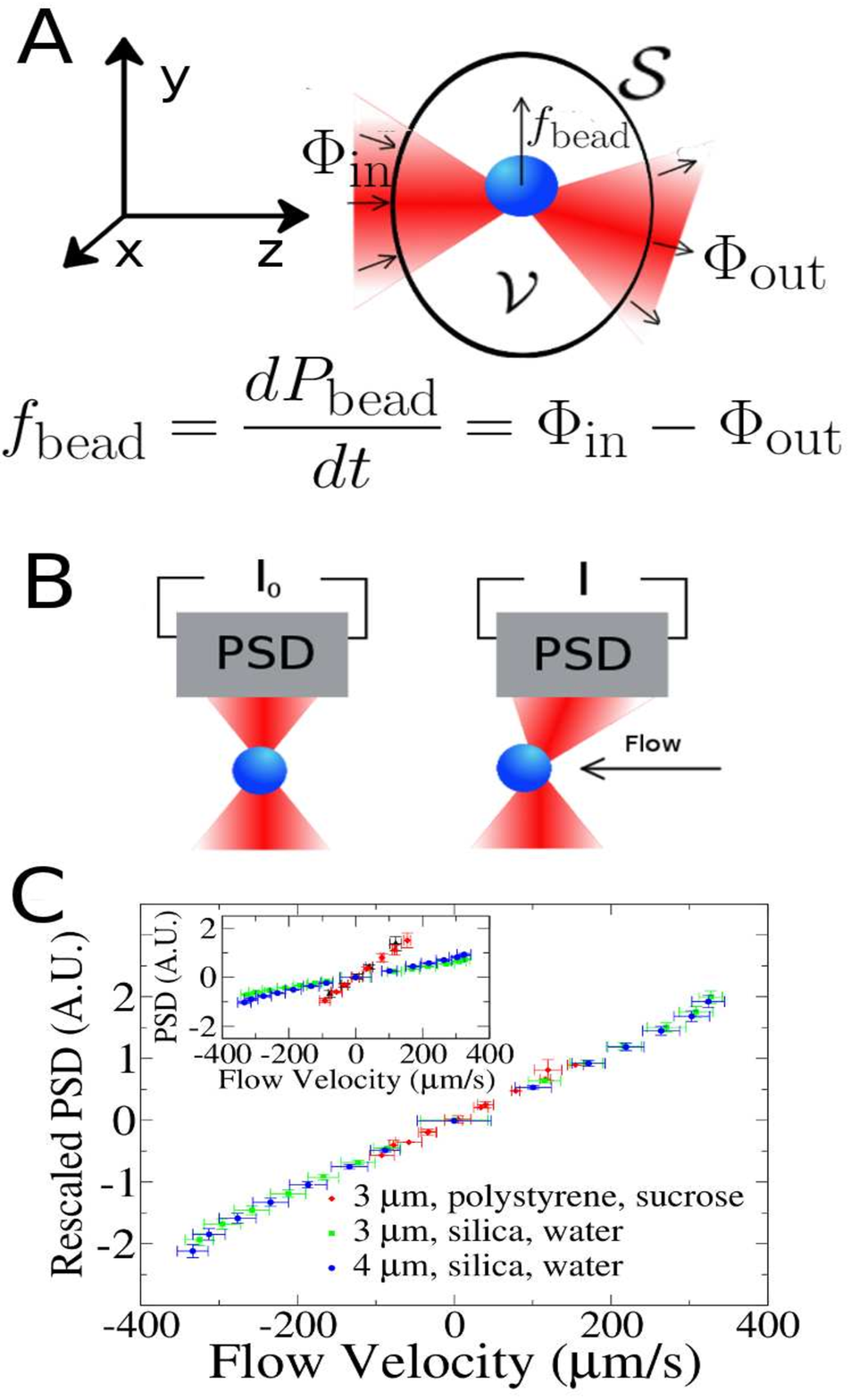}
\caption{\label{Fig:califorcmet}{\bf Force measurement and calibration.} A) The force measurement method, based on linear momentum conservation, exploits the equality between the change 
in total momentum contained in a volume $\mathcal{V}$ enclosed by a surface $\mathcal{S}$ 
and the momentum flux through the surface. The total momentum change inside $\mathcal{V}$ equals to the force acting on the bead. The PSD measures the outgoing flux ($\Phi_\text{out}$) and the ingoing flux ($\Phi_\text{in}$) is determined
 from measurements at zero force with a bead captured in the trap.
B) The PSD returns a current $I$ proportional to the outgoing flux. The difference between the current measured at zero force and the
current measured at a given time is proportional to the instantaneous force, $F_i=\lambda\left(I_i-I_i^0\right),\, i=x,y$. C) PSD response during a Stokes test. Inset: Stokes tests
on beads of different size and materials and different buffer solutions. Results from five different beads were averaged in each case (error bars obtained as rmsd). Different responses are obtained because of different drag forces.
Main plot: when the response is rescaled by the viscosity $\eta$ and the bead radius $r$ all the response curves collapse and the same
calibration factor is obtained under different experimental conditions.} 
\end{figure}
\begin{equation}
f_x=\lambda_x \left(I_x-I_x^0\right); \qquad f_y=\lambda_y \left(I_y-I_y^0\right),
\end{equation}
where $I^0_x,\,I^0_y$ are the currents measured when no force is acting on the trapped bead (Fig. \ref{Fig:califorcmet} B).
This can be done by applying a known force on the trapped bead and measuring the PSD response.
A practical way of doing this is using Stokes law for the drag force on a spherical bead in a viscous flow, $f_\text{Stokes}=6\pi\eta r v$, where $r$ is the radius of the bead, 
$\eta$ the viscosity of the medium and $v$ the flow velocity. The flow is generated by moving the microfluidics chamber with respect to the optical trap at
constant speed $v$ either along $x$ or $y$. The conversion factors $\lambda_x,\lambda_y$ are then measured as: $\lambda_i=f_{\text{Stokes}}/\left(I_i-I^0_i\right)$. 
As we emphasized in the introduction the most appealing feature of force measurement methods based on light momentum conservation is the robustness of calibration. 
Conversion factors are determined by the optical setup and the responsivity of the detector, but they are independent of the details of the experimental setup, such as the index of 
refraction of the trapped bead, its size or shape, the refractive 
index of the fluid medium and laser power. To prove that this property holds in our setup we measured the PSD response (in arbitrary units, AU) 
along the direction of the flow using two different buffer solutions and 
beads of different materials and sizes. One set of experiments was performed using water as fluid medium and silica beads (3 and 4 $\mu m$, Kisker Biotech), the other set was performed using a high concentration
solution of sucrose and polystyrene beads (3$\mu m$, Spherotech). 
Fig. \ref{Fig:califorcmet} C shows the relation between the flow velocity and
the PSD signal along the direction of the flow. Different fluids and different bead sizes lead to different results (Fig. \ref{Fig:califorcmet} C, inset) because both the bead radius and the 
fluid viscosity influence the drag force. Nevertheless if we normalize the PSD signal by the product $r\eta$ the different curves collapse (Fig. \ref{Fig:califorcmet} C, main plot), showing 
that the calibration factor is the same in all cases. However, trap stiffness and maximum trapping force do depend on the refractive indices of bead and medium. 
The typical performance of the trap, when using silica beads in water at 50mW of laser power per bead in the sample region, is shown in Fig. \ref{Fig:ref}.
There we show the force exerted by the trap along the y--axis, as a function of the distance between the center of the trap and the bead. 
To obtain this curve a silica bead is captured on the tip of a micro--pipette by air suction. A trap is then formed at the center of the bead and moved \cite{PNAS.huguet.2010}: when the trap is not focused at the center of the bead, the light
exerts a force on the immobilized bead, which can be measured through the PSD.  
The results concerning the shape of the trap are  summarized in Fig. \ref{Fig:ref}.  
The trap has a narrow linear zone which spans the first few ($\simeq 5$) pN of  applied force and shows a strong non linearity thereafter.
The corresponding trap stiffness is plotted in the lower left inset of Fig. \ref{Fig:ref} as a function of the applied force. The stiffness was obtained as the 
numerical derivative of the force--displacement curve in the main plot.
The non--linear trap stiffness is due to the spherical aberration of the bead acting as a lens on the laser beams. Such effect is important in optical tweezers
when the radius of the trapped bead is larger than the focal spot. Nonlinearity could be reduced using smaller beads but in this case the maximum trapping force
would be much smaller.

\begin{figure}
 \centering
\includegraphics[width=0.99\textwidth]{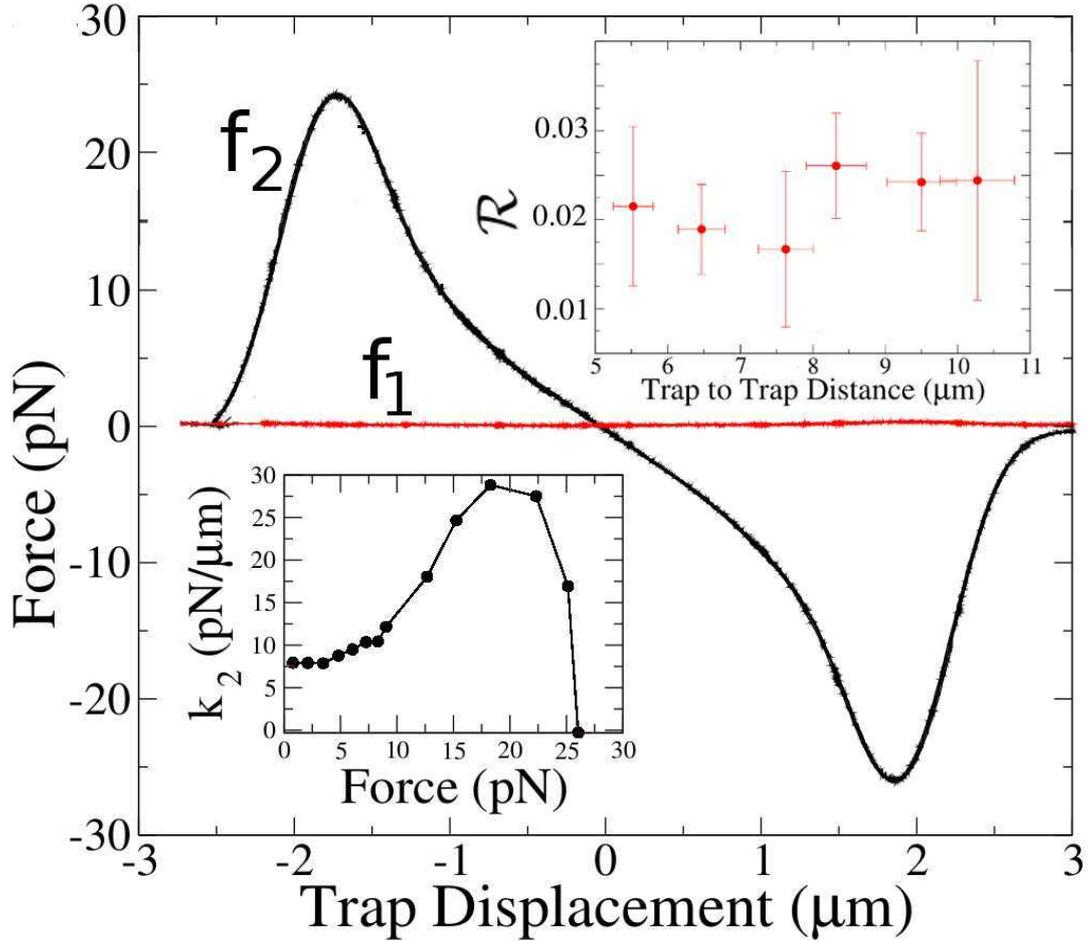}
\caption{\label{Fig:ref}{\bf Trap shape and reflection effects in the DT setup.} 
A) Main plot: force exerted by the laser beam on the bead 
as a function of the displacement of the bead from the center of the trap 2 (dark line). The measurement was done by immobilizing a bead on the tip of a micropipette
(see text).
The fair line shows the spurious force measured in the second empty trap 
which is due to reflected light ($f\leq 0.2$ pN).  The lower left inset shows the local stiffness of the optical trap, as
a function of the force, as obtained from a numerical derivative of the curve in the main plot. Trap stiffness varies
from 10 pN/$\mu$m at low forces to a maximum of 25 pN/$\mu$m, due to strong nonlinear effects. In these measurements the bead is not free
to move along the z axis.
The upper right inset shows the reflectivity parameter $\mathcal{R}$ (as defined in text) as a function of the trap--to--trap distance. The measured reflectivity is never above
3\% and can be neglected in our force measurements. 
}
\end{figure}

\section{On the magnitude of reflection effects at low forces}\label{fff}

This experimental setup uses counter-propagating beams from different sources. It is thus free from cross-talk effects between the beams as
described in \cite{RevSciInst.mangeol.2008}. Nevertheless a systematic source of error in the measurements can arise due to reflection effects.
In the ST configuration reflection is not a source of error. Reflected light creates an actual force on the bead and the Minitweezer optics collects
and scores reflected light in the correct way to account for reflection forces. On the contrary in the DT setup the reflected light from one bead
adds to the wrong PSD. In fact when the light of one beam, say beam 1, hits the surface of a trapped bead, part of the incoming radiation will be transmitted and part reflected. 
Light reflected from beam 1 can then propagate backwards along the optical axis, reaching the PSD which is meant to measure the force acting on beam 2 (PSD-2).
As a consequence, the signal reaching PSD-1 is composed of light from beam 1 which has been transmitted and light from beam 2 which 
has been reflected and vice-versa for PSD-1. (This is not a problem in the ST setup as in that case one is only concerned with the total deflection
of the two beams.) In terms of the measured forces, let $f^T_i,f^R_i$ denote the measured signals due to transmitted and reflected light from laser $i$ respectively. 
If $f_i$ is equal to the net force signal measured in PSD-$i$, then we have: 
\begin{equation}\label{refle}
\begin{split}
f_1=f^T_1+f^R_2\\
f_2=f^T_2+f^R_1.
\end{split}
\end{equation}
The magnitude of the reflection effect can be quantified: when two beads are optically trapped in the same fluid at rest, their static fluctuations are independent \cite{Prl.meiners.1999}. 
The independence of the static fluctuations is expressed mathematically by saying that the covariance between the signals coming from the two traps is equal to zero.
Eq. \eqref{refle} can be used to show how, due to reflection effects, a spurious covariance can arise, even in absence of a true physical interaction between the two beads. The covariance is given by:
\begin{equation}
 \langle f_1 f_2\rangle=\langle f^T_1 f^T_2 \rangle+\langle f^T_2 f^R_2 \rangle+\langle f^T_1 f^R_1 \rangle+\langle f^R_1 f^R_2 \rangle=
\langle f^T_2 f^R_2 \rangle+\langle f^T_1 f^R_1 \rangle,
\end{equation}
where $\langle \cdots \rangle$ denotes thermal average. Note that in the last equality we used the independence of force fluctuations 
in the two beads and $\langle f^T_i\rangle=\langle f^R_i\rangle=0$. 
A relative measure of reflection effects is given by the reflectivity parameter $\mathcal{R}$
\begin{equation}\label{relref}
 \mathcal{R}=\frac{\langle f^T_2 f^R_2 \rangle+\langle f^T_1 f^R_1 \rangle}{\langle \left(f^T_1\right)^2\rangle+\langle \left(f^T_2\right)^2\rangle}\simeq
\frac{\langle f_1 f_2\rangle}{\langle \left(f_1\right)^2\rangle+\langle \left(f_2\right)^2\rangle},
\end{equation}
where we take $\langle (f^{T}_i)^{2}\rangle\simeq \langle (f_i)^{2}\rangle$ since $\langle f^{T}_i\rangle \gg\langle f^{R}_i\rangle $.
We have measured $\mathcal{R}$ using pairs of silica beads trapped in water and recorded the two force signals at 50 kHz acquisition bandwidth. 
Such measurements were repeated at different trap--to--trap distances. In every case reflection effects proved less than $3\%$ (see Fig. \ref{Fig:ref} upper right inset).

\section{Fluctuation analysis in DT setups: resolution and stiffness measurements}\label{sectionfluct}

\subsection{Fluctuation analysis: resolution limits}\label{rslim}

\begin{center}
\begin{figure}
\centering
\includegraphics[width=0.6\textwidth]{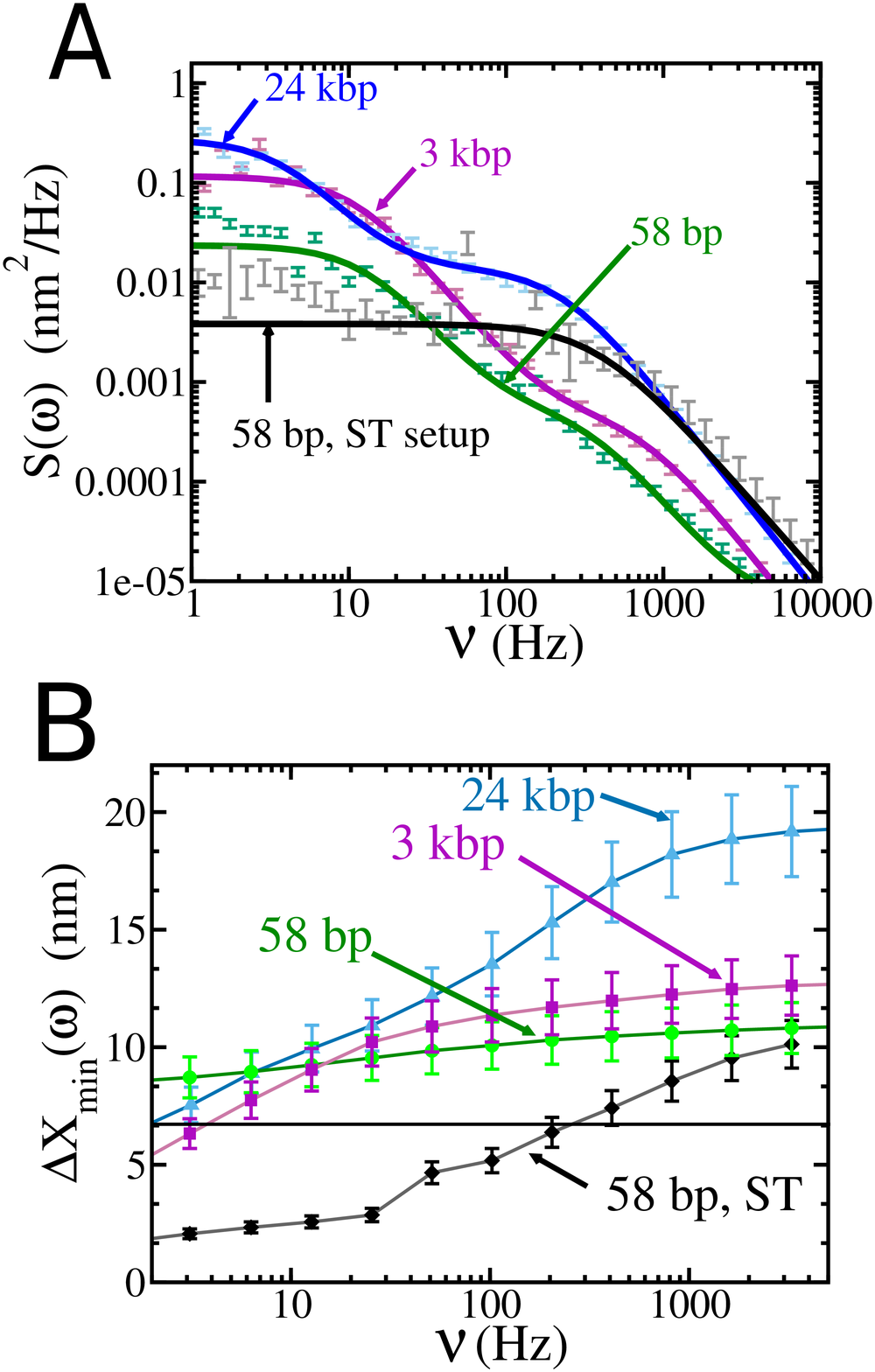}
\caption{\label{Fig:vars} {\bf Fluctuation analysis.} A) the fluctuation spectrum of the differential coordinate obtained in the DT setup on three tethers of different lengths under 10 pN tension:
a 24kbp tether, a 3kbp tether and a 58 bp tether with an inserted hairpin. Data were fit to the sum of two Lorentzian (Eq. \eqref{DL}). Fit results are shown in Table \ref{Table:DL} . One Lorentzian arises due
to fluctuations along the pulling direction, while the other arises due to transverse fluctuations due to misalignment \cite{BiophysJ.ribezzi.2012}. Transverse
fluctuations decay over longer characteristic timescales ($\nu_\text{slow}^{-1}$), their contribution being important at smaller frequencies. The total area covered by the
power spectrum (the full variance of the signal) is seen to decrease with the tether's length. B) resolution of the instrument as a function of bandwidth
for different tethers \eqref{resdef}. Due to the large amplitude of the low frequency component in shorter tethers, in the DT setup the minimum resolvable length change is almost insensitive to averaging.}
\end{figure}
\end{center}

We will now discuss the noise level in our setup and the different noise sources. We shall at first consider the ideal case in which the tether is perfectly aligned along 
the pulling direction in the infinite bandwidth limit to later introduce averaging and misalignment effects \cite{BiophysJ.ribezzi.2012}.
It is well known that the resolution of DT setups is not limited by the trap stiffness \cite{PNAS.moffitt.2006}. The two force signals coming from the traps can be linearly
combined and the resolution is set by that combination which displays the least variance, the so-called differential signal. If the DT setup is symmetric, i.e. the traps have equal stiffnesses, the differential 
signal is simply given by the difference of the signals coming from both traps. If the traps are asymmetric, with stiffnesses $k_1$ and $k_2$, a minimum in the variance 
of the linear combination:
\begin{equation}\label{defif}
f_\phi=\phi f_1-(1-\phi)f_2,
\end{equation}
can still be found. The variance of $f_\phi$ is:
\begin{equation}\label{parab}
\begin{split}
&\sigma_{\phi}^2=\langle f_\phi^2\rangle-\langle f_\phi\rangle^2=\\
&=k_BT\left(\frac{(k_2)^2(k_1+k_m)}{k_2 k_m+k_1(k_2+k_m)}-2k_2\phi+(k_1+k_2)\phi^2\right),
\end{split}
\end{equation}
and the minimum is found for 
\begin{equation}\label{opttt}
\phi^*=\frac{k_2}{k_1+k_2}\qquad.
\end{equation}
These results are valid in the ideal case in which the tether is perfectly aligned along the pulling direction and there are no instrumental noise sources, such as electronic
or ambient noise, which are nevertheless always present. Misalignment \cite{BiophysJ.ribezzi.2012} is a major source of noise in our setup and does actually set the resolution
limit. The different noise contributions are easily identified by looking at the noise power spectrum $S(\nu)$. 
In Fig \ref{Fig:vars}A we show the power spectrum of fluctuations in the differential coordinate using three different dsDNA tethers of different lengths: 24kbps, 3kbps and 58bps \cite{suppmatt2}.
The force fluctuations spectra were converted to distance fluctuations using the trap stiffness (0.02 pN/nm in the DT setup, 0.06 pN/nm in the ST setup), so that
setups with different trap stiffnesses can be compared.
 The three power spectra obtained in the DT setup can be fitted with a double Lorentzian behavior:
 \begin{equation}\label{DL}
 S_\text{fit}(\nu)=\frac{A_\text{slow}}{\nu_\text{slow}^2+\nu^2}+\frac{A_\text{fast}}{\nu_\text{fast}^2+\nu^2}
 \end{equation}
The fast contribution is due to fluctuations along the pulling direction, while the slow contribution is due to misalignment and fluctuations along
the optical axis. Figure \ref{Fig:vars}A also shows the power spectrum obtained on the 58 bp tether in the ST setup. A comparison of the two spectra for the 58 bp shows that, 
although the total area of the power spectrum is comparable in the two setups, their frequency distribution is different. 
At high frequencies the power spectrum is larger in the ST setup than in the DT setup, while the contrary is true at low frequencies.
The power spectrum can be used to define a frequency-dependent variance, which describes the expected behavior of the noise under averaging:
\begin{equation}\label{sw}
\sigma_\phi^2(\nu)=2\int_0^\nu S(\nu')d\nu'.
\end{equation}

In equilibrium experiments, the mean force in the two traps is the same $\langle f_1\rangle=-\langle f_2\rangle=\langle f\rangle$, so that the mean
of \eqref{defif} is independent of $\phi$ (within force calibration errors, $\simeq$ 3\%, see Fig \ref{Fig:DNAstiff} B):
\begin{equation}\label{mean}
\langle f_\phi\rangle=\phi \langle f_1\rangle -(1-\phi)\langle f_2\rangle=f\qquad .
\end{equation}
Imagine now a biochemical process changing the tether's length by $\Delta x$ and thus the mean force by $\Delta f$. This process will be observable
if the force change is at least twice as big as the rmsd of the force differential signal (cf. Eq \eqref{sw}): 
\begin{equation}\label{resolve}
 \frac{\Delta f}{2\sigma_\phi(\nu)}>1.
\end{equation}
This can be easily translated to a requirement on $\Delta x$ as:
\begin{equation}
\frac{k_T}{2}\frac{\Delta x}{2\sigma_\phi(\nu)}>1,
\end{equation}
where we used $\Delta f=k_T\frac{\Delta x}2$. We thus define the minimum resolvable length change as:
\begin{equation}\label{resdef}
\Delta X_\text{min}(\nu)=\frac{4\sigma_\phi(\nu)}{k_T},
\end{equation}
with $k_T$ the typical trap stiffness. In Fig \ref{Fig:vars}B we plot $\Delta X_\text{min}(\nu)$, 
for tethers of three different lengths: 24kbs, 3 kbp and 58 bps. 
Clearly, the resolution at high bandwidth is better for the shorter tethers which display smaller longitudinal fluctuations. Due to the presence of the slow
component, in the DT setup, $\Delta X_\text{min}(\nu)$ is almost insensitive to box averaging. This is not true for the ST setup, which is less prone to misalignment. 
Slow fluctuations account for most of the noise in the DT setup in the cases of 3 kbps and 58 bps. In this sense we can say that misalignment
sets the limit to the resolution of the instrument.
\begin{center}
\begin{table}
\centering
\begin{tabular}{|c|c|c|c|c|}
\hline \hline
 & $A_\text{slow}$ (nm$^2$Hz) & $\nu_\text{slow}$ (Hz) & $A_\text{fast}$ (nm$^2$Hz)& $\nu_\text{fast}$ (Hz)\\
24 kbp, DT setup & $3.8\pm0.8$ & $3.7\pm0.5$ & $710\pm40$ & $230\pm20$\\
3 kbp, DT setup & $15\pm2$ & $11\pm1$ & $271\pm30$ & $850\pm100$ \\
58 bp, DT setup & $15\pm4$ & $11\pm2$ & $71\pm16$ & $420\pm100$ \\
58 bp, ST setup & - & - & $1200\pm150$ & $550\pm50$\\
\hline
\end{tabular}
\caption{\label{Table:DL} Results of the double Lorentzian fits (Eq. \eqref{DL}) to the measured spectra in Fig.\ref{Fig:vars}. $A_\text{slow}$ and $A_\text{fast}$
are the amplitude of the slow and fast component respectively, while $\nu_\text{slow}$ and $\nu_\text{fast}$ are the corresponding corner frequencies.
It might look surprising that the corner frequency is higher for the 3kbp tether than for the 58 bases tether as the latter is stiffer. This is due to the
fact that hydrodynamic interactions are much bigger in the case of the shortest tether.}
\end{table}
\end{center}

\subsection{Fluctuation analysis: stiffness measurements}\label{fonono}

\begin{figure}
\centering\includegraphics[width=0.8\textwidth]{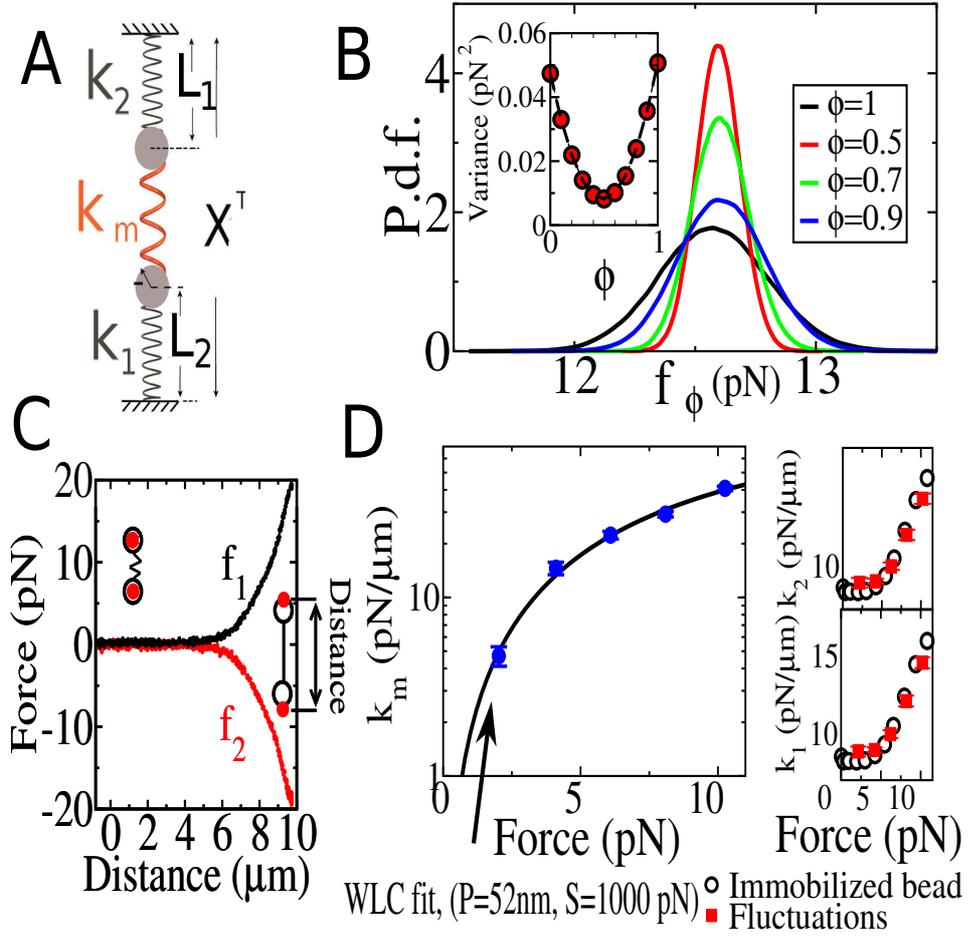} 
\caption{\label{Fig:DNAstiff}{\bf Elasticity of dsDNA tethers.} A) Linear model of the dumbbell shown in Fig. \ref{Fig:1} A, where three elastic elements with different 
stiffnesses are arranged in series:
Trap 1 ($k_1$), Trap 2 ($k_2$) and the tether ($k_m$).
B) Probability distribution and variance of the generalized force signal, $f_\phi$, defined in Eq. \eqref{defif}. Note that the slight shift in the value of the mean force shown 
in the main plot is due to small force calibration errors ($\simeq$ 3\%). In the inset we show the variance computed from the probability distribution at different values of $\phi$ (solid symbols) and
 the parabolic fit used to measure the stiffness of both traps and the tether through Eq. \eqref{parab} (dashed line). 
C) Force--Distance curves ($f_1,f_2$) for dsDNA half-$\lambda$ tethers
measured in the DT setup. Note that the two forces have equal averages and opposite sign. Data for two different molecules are shown.
D) Main plot: molecular stiffness ($k_m$) measured for 
5 different molecules. The continuous line shows a fit to the WLC model \eqref{kWLC}, giving $P=52\pm 4$ nm, $S=1000\pm200$ pN. 
In the smaller plots we compare the stiffness values of the two traps, $k_1$ and $k_2$,
measured through Eq. \eqref{parab} (open symbols) with those measured by immobilizing the bead on the micropipette (solid symbols), see Section \ref{cali} . 
Measurements agree within experimental errors.}
\end{figure}

In equilibrium experiments, i.e. at fixed trap--to--trap distance the dumbbell in Fig. \ref{Fig:1} A can be thought of as the series of three linear elastic elements (Fig. \ref{Fig:DNAstiff} A).
In this approximation a recently developed method \cite{BiophysJ.ribezzi.2012} links the covariance of the measured force signals, $\sigma^2_{ij}=\langle f_if_j\rangle-\langle f_i\rangle \langle f_j\rangle,\,\, i=1,2$,
to  the stiffnesses of both traps $k_1,k_2$ and tether, $k_m$.
\begin{eqnarray}
 k_1&=&\frac{\sigma_{11}^2+\sigma_{12}^2}{k_B T}\label{ia}\\
k_2&=&\frac{\sigma_{22}^2+\sigma_{12}^2}{k_BT}\label{ib}\\
k_m&=&\frac{1}{k_BT}\frac{\sigma_{12}^2\left( \sigma_{11}^2+\sigma_{12}^2\right)\left(  \sigma_{22}^2+\sigma_{12}^2\right)} { \sigma_{11}^2\sigma_{22}^2-\sigma_{12}^4}\label{ic}.
\end{eqnarray}
where $k_B$ is the Boltzmann constant and $T$ the absolute temperature.
Our traps are highly nonlinear (Fig. \ref{Fig:ref})  and a practical method to measure the force dependence of trap stiffness proves thus very useful. 

Equations \eqref{ia},\eqref{ib} and \eqref{ic} were applied to measure the elastic response of a 24 kbp dsDNA tether (half of the $\lambda$--phage genome).
The nonlinear elasticity of dsDNA has been studied with OT since two decades \cite{Science.bustamante.1994,Science.smith.1996,BiophysJ.wang.1997,BiophysJ.bouchiat1999}, but it is still attracting much interest, especially as far as short ($50-500$ bp) molecules 
are concerned \cite{BiophysJ.seol.2007,Pre.chen.2009}. The results on DNA elasticity obtained by our method can be compared to the large existing literature on the subject, while the stiffness measurements on the traps can be compared with values obtained by the micro--pipette method
described at the end of Section \ref{cali}. Experiments were performed measuring equilibrium fluctuations in the dumbbell shown in Fig. \ref{Fig:1} A, using 4$\mu$m silica beads
in a Phosphate Buffer Saline solution, NaCl 1M, containing 1 mg/ml of Bovine Serum Albumin (BSA) to passivate glass surfaces. Fluctuation traces were measured at 50 kHz acquisition bandwidth for periods of 10 seconds. The values of the model 
parameters ($k_1,k_2,k_m$) were obtained through Eqs. \eqref{ia},\eqref{ib},\eqref{ic}.  In the inset of Figure \ref{Fig:DNAstiff} B we show the experimental variance of
the generalized force \eqref{defif} as a function of $\phi$ compared to the theoretical result \eqref{parab} using the previously obtained values for $k_1,k_2,k_m$. 
Measurements of the molecular and trap stiffness were performed at different mean forces by changing the trap--to--trap distance (Fig. \ref{Fig:DNAstiff} C). 
Results in \ref{Fig:DNAstiff}B are plotted as a function of the mean force $\langle f_\phi\rangle$.
Above 10 pN the fluctuation spectrum contained a large-amplitude low-frequency component, which was interpreted as coupling of fluctuations along 
the optical axis \cite{BiophysJ.ribezzi.2012}.
The results for $k_m$ were fitted to an extensible Worm--Like--Chain (WLC) model \cite{Macromol.marko.1995}:
\begin{equation}
\ell_\text{WLC}(f)=\ell_0 \left(1-\frac{1}{2}\sqrt{\frac{k_BT}{fP}}+\frac{f}{S}\right), 
\end{equation}
where $\ell_0$ is the tether contour length, $P$ is the persistence length and
$S$ the stretch modulus. This formula is valid for $fP\gg k_BT$.
In terms of the molecular stiffness:
\begin{equation}\label{kWLC}
k_\text{WLC}(f)=\frac{df}{d\ell_\text{WLC}}=\frac 1{\ell_0}\frac 1 {
\sqrt{\frac{k_BT}{16 P}}\left(\frac 1 {f}\right)^{3/2}+\frac{1}{S}}\qquad. 
\end{equation} 
In the fitting procedure the contour length was fixed to $8.2 \mu$m using the crystallographic rise value of $0.34$ nm per base pair, while $P$ and $S$ were varied.
The fit results, $P=47 \pm 4$ nm and $S=1300 \pm 200$ pN, are consistent with those reported in the literature on the elasticity of long dsDNA molecules \cite{Science.bustamante.1994,Science.smith.1996,BiophysJ.wang.1997,BiophysJ.bouchiat1999,Prl.meiners.2000,BiophysJ.seol.2007,Pre.chen.2009}.

\section{Measurements on DNA-hairpins with short handles.}

The stiffness measurements discussed in Section \ref{fonono} involve a long molecule and the bead surfaces are never too close each other during the experiment ($d\geq 2\mu$m).
In the case of short tethers the beads can be almost in contact so that the light scattered from one bead might interact with the other bead before reaching the detector. 
This would result in incorrect force measurement. To check whether the force measurement technique is still working at small bead--to--bead distances ($\leq$ 100 nm) we performed 
experiments on a DNA hairpin with short molecular handles \cite{BiophysJ.Forns.2011}. When tethered by this molecule the beads are only $\simeq 20$ nm apart.

\subsection{Hopping Experiments}\label{mfor}

\begin{figure}
\centering
\includegraphics[width=0.6\textwidth]{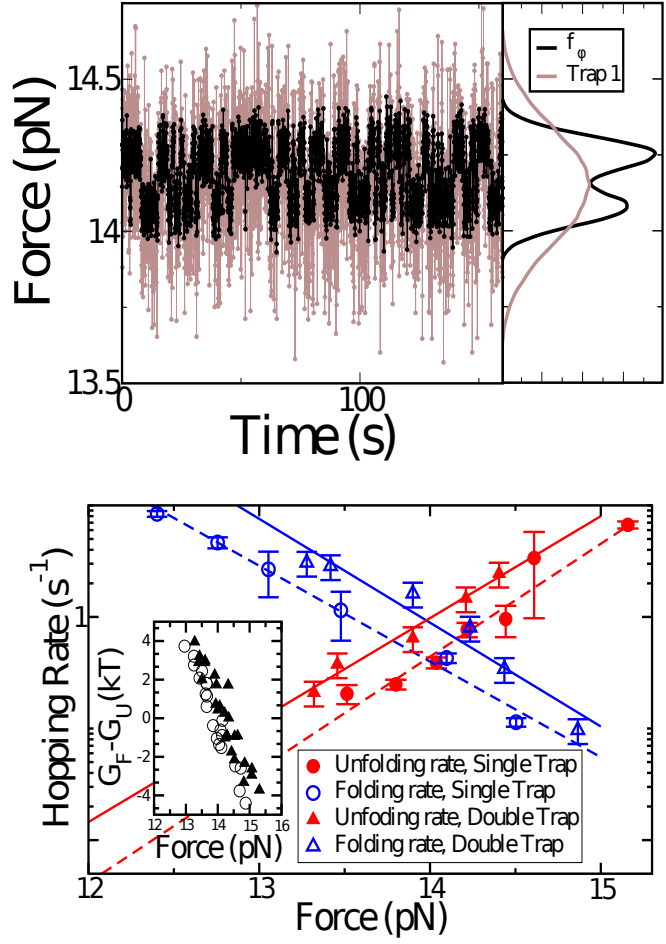} 
\caption{\label{Fig:hairpin}  {\bf Hopping experiments in DNA hairpins.} A) Signal optimization of the signal in the DT setup. The signal coming from one trap (for example $f_1$) has too large
a variance to distinguish between the folded and the unfolded states which do not appear in the force distribution (fair line, right panel). However using $f_{\phi^*}$ we can resolve the two peaks (dark line, right panel). The folding/unfolding transition can be observed by using $f_{\phi^*}$ instead of $f_1$ or $f_2$ (background noisy trace). 
B) Force dependent PM hopping rates of the hairpin measured in two different setups: the DT setup described in this paper (folding/unfolding rates are open/solid triangles) and the ST setup described in \cite{PNAS.huguet.2010} 
(folding/unfolding rates are open/solid circles). Lines are exponential fits to the data using the Bell-Evans model \cite{suppmatt}. Kinetic parameters extracted from the fits are reported in  Table \ref{Table:Kinetic}.
Inset: free energy difference between the folded and the unfolded state as measured
in the two setups via Eq. \eqref{boltzmann} (DT: solid triangles, ST: open circles). Thermodynamic parameters are reported in Table \ref{Table:Thermo}.
}
\end{figure}

One way of studying in detail the force folding/unfolding dynamics of the DNA hairpins, is to perform Passive Mode (PM) hopping experiments.
In these experiments, neither the force nor the molecular extension is kept constant, the control parameter being the trap--to--trap distance ($X_T$ in 
Fig. \ref{Fig:DNAstiff} A).
In PM experiments the average force in the folded and unfolded states are different since the length of the tether changes upon unfolding (folding): 
when the hairpin unfolds (folds) the ssDNA is released (captured) leading to increased (decreased) tether extension. The beads 
move towards (away from) the center of the traps and a force jump is measured. During a hopping experiment a molecule performs several folding/unfolding cycles (see Fig. \ref{Fig:hairpin} A), whose kinetic rates can be modulated by changing the trap--to--trap distance and thus the average forces.
The dwell force distribution of a hopping experiment trace shows two different peaks corresponding to the folded and unfolded states.
These peaks are broadened by thermal fluctuations to an extent which depends on the trap stiffness. The Signal--to--Noise Ratio (SNR) can be defined in this context as:
\begin{equation}\label{snr}
\text{SNR(1 kHz)}=\frac{|f_F-f_U|}{2\sigma_\phi(\text{1 kHz})}\qquad,
\end{equation}
where $f_F,f_U$ are the forces at which the peaks are located and  $\sigma_\phi(\nu)$ the width of the peaks at bandwidth $\nu$ (Eq. \eqref{sw}).
In a DT setup the SNR can be enhanced using the generalized force (Eq. \eqref{defif}).
The ``optimal signal" for hopping experiments, i.e. the signal with highest SNR is given by the same value $\phi^*$, Eq. \eqref{opttt}. Indeed $f_{\phi^*}$ minimizes the variance of force fluctuations $\sigma_{\phi}^2$ (Eq. \ref{parab}),
and thus the denominator of Eq.\eqref{snr}. In contrast, the numerator of \eqref{snr}, which only depends on mean values, is independent of $\phi$.  
In our experimental conditions the best signal is always found for $\phi^*\simeq \frac{1}{2}$, but the actual value can slightly differ due to asymmetries in beads 
and traps. Experimental $\phi^*$ values for the experiments reported in this study vary in the range $0.5\pm 0.1$.
The optimal hopping signal can be used to precisely measure the relative population of the two states by fitting the force dwell distribution to the sum of two Gaussians.
By Boltzmann formula we have:
\begin{equation}\label{boltzmann}
-\beta^{-1}\log\left(\frac{P_F}{P_U}\right)=G_U-G_F=-\Delta G_0+x_m\left(\frac{f_U+f_F}{2}\right),\\
\end{equation}
where $G_U$ and $G_F$ are the Free energies of the folded and of the unfolded state respectively, $\Delta G_0$ is the free energy difference at zero force and
$x_m$ is the change in molecular extension upon unfolding. A linear fit to the force dependence of the l.h.s. of Eq.\eqref{boltzmann} determines $\Delta G_0$ and  $x_m$.
The thermodynamics of PM experiments and the derivation of Eq. \eqref{boltzmann} can be found in \cite{JSTAT.mossa.2009,JSTAT.manosas.2009}.
In addition, the inverse of the average lifetimes yield the kinetic rates at different trap positions. The force dependence of the rates can be interpreted using Bell-Evans
theory to infer the free energy difference at zero force $\Delta G_0$, the distance from the folded (unfolded) state to the transition state $x_{FU}$ ($x_{UF}$) and
the co-existence kinetic rate $k_c$ \cite{suppmatt}. 
Figure \ref{Fig:hairpin} C compares the results of PM hopping experiments performed using both the DT and the ST setups.
Thermodynamic quantities, measured either analyzing the relative population of the two states (Eq. \eqref{boltzmann}) or using detailed balance, are largely independent of the experimental setup
being used and they are consistent within the experimental error. 
Kinetics does instead differ in the two setups: the apparent coexistence rate is more than two times higher for DT setup than it is for ST setup, in agreement
with previous findings on the kinetic rates for tethers of different lengths and traps of different stiffness \cite{Prl.greenleaf.2005},\cite{BiophysJ.Forns.2011}.
The consistency of the thermodynamic results shows that even when the beads are very close, forces are still correctly measured in the DT setup.

\begin{center}
\begin{table}
\centering
\begin{tabular}{|c|c|c|c|c|c|c|}
\hline \hline
&  $ x_\text{FU}(nm)$ & $ x_\text{UF}(nm) $ & $x_m (nm)$ & $\Delta G(k_BT)$  & $f_c(pN)$ & $k_c(s^{-1})$ \\
\hline
ST & $9.5\pm0.5$ & $8.1\pm0.5 $ & $17.7\pm0.7$ & $60\pm3$ &  $14\pm 1$ & $0.4\pm0.3$ \\
\hline
DT & $8.6\pm0.7$ & $8.8\pm0.7 $ & $17.5\pm0.9$ & $59\pm 4$ &$ 14\pm1$ & $0.9\pm0.4$\\
\hline
\end{tabular}
\caption{\label{Table:Kinetic} Kinetic parameters obtained by fitting Bell--Evans model rates (\cite{suppmatt}) to data in Fig. \ref{Fig:hairpin} C.
The co-existence kinetic rate $k_c$ depends on the setup, but thermodynamic quantities and free energy landscape parameters,
$x_\text{FU}, x_\text{UF}, x_m, \Delta G, f_c$, do not. The results are
averaged over 5 different molecules and errors are standard error over different molecules.}
\end{table}
\end{center}

\begin{center}
\begin{table}[ht]
\centering
\begin{tabular}{|c|c|c|c|}
 \hline \hline
&  $ x_m(nm)$ & $\Delta G_0(k_BT)$  & $f_c(pN)$ \\
 \hline
   ST & $18\pm1$ & $60\pm3$ &  $14\pm 1$ \\
     \hline
       DT & $17\pm1$ & $58\pm 5$ &$ 14\pm1$\\
         \hline
	 \end{tabular}
\caption{\label{Table:Thermo} Thermodynamic parameters of the hairpin obtained by fitting Eq. \eqref{boltzmann} to the data in the inset of Fig. \ref{Fig:hairpin} C. 
The results are consistent within experimental error. 
The results are averaged over 5 different molecules. Errors are standard error over measured over different molecules.}
	 \end{table}
	 \end{center}

\subsection{Elastic fluctuations}

\begin{figure}
\centering
\includegraphics[width=0.7\textwidth]{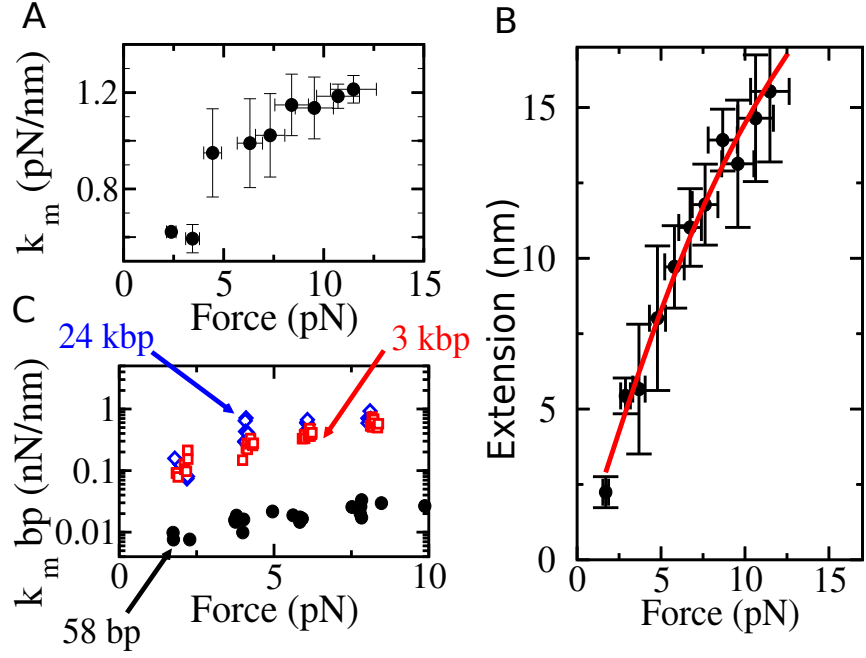} 
\caption{\label{Fig:hairpinela}  {\bf Stiffness measurements on short tethers.} A) Stifness of a molecular construct consisting of two 29bp handles interspaced by
a molecular hairpin. The stiffness was measured from fluctuations, removing the contribution due to misalignment, as detailed in \cite{BiophysJ.ribezzi.2012}.
The hairpin stays closed in the force range explored. B) Data points show the Force Extension curve obtained form the molecular stiffness by integration (Eq. \eqref{fakefec}),
the continuous line shows a freely-jointed chain fit to the data. Fit parameters $b=1\pm0.1$ nm $S=20\pm2$ pN. C) Comparison of measured stiffness per basepair on dsDNA tethers
of different length. Open symbols: data for a 24kbp (diamonds) and 3kb (squares) tethers. Solid symbols data for the 58 bp tether. Stiffness per basepair in
the shortest tether is an order of magnitude smaller than in longer tethers. Data obtained from 5 different molecules are shown in the three cases.}
\end{figure}

Although the elasticity of dsDNA is well established for kilobase long tethers, several recent experiments \cite{BiophysJ.seol.2007,Pre.chen.2009,BiophysJ.Forns.2011} and atomistic simulations \cite{Prl.noy.2012} argue that DNA could be
much more flexible at shorter length scales. In Section \ref{mfor} we have compared results obtained in the DT and ST 
setup to prove that force is correctly measured when the beads are very close to each other. This feature can be exploited to study the elasticity of very short
tethers: a construct formed by 29 bps handles interspaced by a molecular hairpin. In this case the measurement of the FEC, as performed in Section \ref{fonono} is not possible.
On the one hand the large difference in stiffness between the traps ($\simeq 0.02$ pN/nm) and the tether ($\simeq 1$ pN/nm) makes it difficult to measure the
molecular extension, $x_m$. On the other hand misalignment will strongly affect measurements on such short tethers \cite{BiophysJ.ribezzi.2012} and must be taken into account in data analysis (while such
effect was disregarded in Section \ref{fonono}). The molecular stiffness $k_m$ can be obtained from fluctuation measurements after removing misalignment \cite{BiophysJ.ribezzi.2012}, and the FEC thus obtained
by integration:
\begin{equation}\label{fakefec}
x_m(f)=\int_0^f df' \frac1{k_m(f')}\qquad .
\end{equation}
Stiffness measurements are shown in Fig. \ref{Fig:hairpinela} A, while the reconstructed FEC is shown in Fig. \ref{Fig:hairpinela} B. For such short tether the approximation used in Section \ref{fonono} ($Pf\gg k_BT$) for the force extension
curve is not suitable anymore. The elasticity of these short handles follows an extensible freely-jointed chain behavior \cite{Science.smith.1996} (Fig. \ref{Fig:hairpinela} B, continuous curve):
\begin{equation}\label{ffjjcc}
\ell_\text{FJC}(f)=\ell_0\left(\coth\left(\frac{bf}{k_BT}\right)-\frac{k_bT}{bf}\right)\left(1+\frac{f}{S}\right).
\end{equation}
Here $\ell_0$ is the contour length, $b$ is the Kuhn length and $S$ the stretch modulus. Fitting Eq. \ref{ffjjcc} to the data gives $b=1\pm0.1$ nm and $S=20\pm2$ pN.
The tether appears much softer than what it would be expected if the elastic parameters valid for kilobase sized tethers are extrapolated to these short length scales.
This is made evident in Fig. \ref{Fig:hairpinela} C comparing the stiffness per bp as obtained on tethers of different length. Stiffness per basepair is approximately constant for long tethers (24 kbp and 3 kbp, open symbols in the figure), 
while it is an order of magnitude smaller for the  shortest tether (58 bp, solid symbols). 

\section{Discussion and Conclusions}

In this paper we report a novel dual--trap optical tweezers design that uses counter-propagating beams. This setup can be used to perform single--molecule manipulation,
with most of the known advantages of all--optical setups, although lacking drift compensation between the two traps which characterizes co-propagating tweezers. In addition the new setup has the following features:
\begin{enumerate}
 \item {\bf Direct force measurement.} The counter-propagating design offers the possibility to clearly separate the light coming from each trap, 
thus allowing to measure changes in light momentum in each beam separately, and therefore measure the force in each trap. The only source of mixing between the two 
light beams arises as a consequence of reflection effects by the trapped beads but these have been proven to be smaller than 3\%. 

\item {\bf Force measurements on short tethers.}  The setup has been shown to correctly measure forces, even when the beads are 
very close, down to $\simeq$ 20 nm. Such measurements have not been reported in co--propagating setups. A comparison of the performance the two geometries on these tethers would be interesting.

\item {\bf Low power.} The setup uses low laser power (50 mW per beam). Although this limits the trap stiffness it also reduces the heating of the sample region, 
                       the tether damage from Reactive Oxygen Singlets \cite{BiophysJ.landry.2009} and
                       possible optical binding effects \cite{Prl.burns.1989}.  

\item{\bf Versatility.} The setup described here shares the same optical design of the single--trap setup in \cite{PNAS.huguet.2010}, it is thus possible to switch from 
one to the other, according to the experimental situation, without even requiring a new calibration.
\end{enumerate}
As any other design the counter-propagating setup also has some drawbacks:

\begin{enumerate}
\item{\bf Low trap stiffness.} The low (per Watt) trap stiffness is a consequence of the design of the instrument: in order to measure forces correctly all the light must be collected after it
interacts with the trapped object and this sets an upper bound on the maximum numerical aperture (NA) of the beams. Nevertheless the low trap stiffness is not by itself a limit
to the resolution of the DT setup. 

\item{\bf Misalignment.} Misalignment is one of the problems which affect the counter-propagating design. 
It has been shown to affect the fluctuation spectrum (Section \ref{sectionfluct}) and 
to limit the resolution of the setup. Its effects must be taken into account when measuring elastic fluctuations \cite{BiophysJ.ribezzi.2012}, but can be neglected in pulling
or hopping experiments.

\end{enumerate}

We have developed a methodology to extract the stiffness of the optical traps and the tether based on measurements of force fluctuations in each trap. 
This method is based on the analysis of a generalized force signal $f_\phi$ (cf. Eq. \eqref{defif}) and its fluctuation spectrum, making it possible to evaluate 
the trap yield at different forces and recover the FEC curve of the tether with high fidelity. The greatest advantage of this methodology is the possibility
to measure the stiffness of the optical traps over a  range of forces ($0-10$ pN in our setup) where nonlinear effects are important and the approximation of a linear trap fails. 
Such method is of interest for generic DT setups.
Above 10 pN this methodology can still be applied, however, a correction for the coupling with force fluctuations occurring along the optical axis must 
be included, as detailed in Ref. \cite{BiophysJ.ribezzi.2012}.

A further development of the present work would be the design of a similar double--trap
setup able to reach higher trap stiffnesses. Even if one is not willing to increase the laser power, higher stiffnesses could be achieved 
in several ways. In the first place one could use laser beams with larger beam  waist, which 
would lead to a larger effective numerical aperture and better trapping efficiency along the optical axis, although this could reduce the maximum trapping force. The MiniTweezers setup was initially designed to perform experiments in a single--trap setup where
forces as high as 50 pN per trap are reached (at $\simeq$50 mW laser power). In the DT setup and for comparable laser powers, it is difficult to reach forces above 20 pN, 
and the beam deflection is accordingly smaller. Therefore it should be possible to increase the beam waist and still collect all the deflected light. 
Indeed it has recently been shown that it is possible to use large numerical aperture beams while directly measuring forces \cite{Optexp.farre.2010}, although this requires a more complex instrumental
design.
A second option would be the use of different beam modes as for example the donought-shaped $\text{TEM}^*_{01} $. This idea has been explored theoretically by Ashkin \cite{BiophysJ.ashkin.1992} and
has by now been given different experimental realizations. In this laser mode the intensity profile in the plane perpendicular to the optical axis shows a central dark 
spot surrounded by a bright ring. This should reduce the scattering force along the optical axis, which takes the greatest contribution from light rays at
the center of the beam, and enhance the gradient force by the outer rays. In both cases, whether the beam waist is increased or the beam shape is changed, beads with a higher (but not
too high) effective refractive index could be used leading to better maximum trapping forces and stiffnesses.
The counter-propagating setup discussed in this paper is of direct interest to all labs using counter-propagating ST setups for single molecule experiments, which 
could readily switch to the DT configuration. Moreover the same design could be useful to those willing to implement direct force measurement
in a DT setup in a simple way. 

\section*{Financial Support}

FR is supported by  MICINN FIS2010-19342, HFSP Grant No. RGP55-2008, and ICREA Academia grants. MR and JMH are supported by HFSP Grant No. RGP55-2008.

\section*{Supplementary Information}
\setcounter{section}{0}

\section{Sample preparation}\label{molbio}

The DNA haripins used in the experiments reported in the Main Text have a 20 bp stem and 5 or 8 bases loop. 
This hairpin had the open ends of the stem linked to 29 bp double stranded DNA handles which act as spacers. The free ends 
of the DNA handles were marked with digoxigenin on one end and with biotin at the other end.
The hairpin sequence is schematically represented in Fig. S2.
Syntesis and characterization of similar molecules is described in \cite{BiophysJ.forns.2011}.


For the stiffness measurements he 24kbp ds--DNA was obtained by digestion from the genome of  phage $\lambda$.
Also in this case the ends of the molecule were marked with biotin and digoxigenin
In order to manipulate the molecules these had the two extremes chemically linked to 4 $\mu m$ Silica beads.
Beads (Kisker Biotech) were coated with either antidigoxigenin or streptavidin.
The hydrodynamic measurements were performed on the hairpins, on the 24 kbp ds-DNA and on two other thethers
of 3kbp and 1.2 kbp respectively. These tethers were obtained by digestion and PCR amplification of the phage
$\lambda$ genome.
All experiments were performed in a microfluidics chamber formed by two coverslips interspaced with parafilm. 
The chamber has three channels: a central one where experiments are carried out and two (upper and lower) channels
that are connected to the central one by two dispenser tubes. 
Anti-dig coated beads
were first incubated with the molecule of interest and then introduced in the microfluidics chamber through one of the dispenser tubes. Once the
anti-dig coated bead was trapped a streptavidin coated bead was introduced through a second dispenser tube and trapped in the second trap.
The connection was then formed directly inside the microfluidics chamber. 
All experiments on DNA tethers were performed in PBS buffer solution at pH 7.4, 1M NaCl, at $25^o$ C. 
This buffer solution was found to greatly reduce the nonspecific adsorption of DNA on silica, allowing the use of commercial beads from Kisker Biotech 
without any specific preparation or coating. We dissolved $1mg/\mu l$ Bovine Serum Albumine in the buffer in order to reduce 
silica-silica interactions.

\section{Reconstruction of a coarse free energy landscape}\label{Coarse}

The mechanical folding and unfolding of nucleic acid hairpins is commonly described in terms of a reaction coordinate
and of the corresponding free energy landscape. When subject to force, the end-to-end distance of the molecule along the 
force axis is an adequate reaction coordinate for the folding-unfolding reaction pathway. For a given applied force $f$, a single 
kinetic pathway for the unfolding and folding reactions is usually considered, characterized by
a single transition state (TS). The TS is the highest free--energy state encountered along the reaction coordinate and determines
the kinetics of the folding-unfolding reaction. The model involves four parameters: the free energy of folding at zero force, $\Delta G=G_F-G_U$, the height of the kinetic barrier $B_0$, defined
as the free energy difference between the transition state and the folded (F) state extrapolated to zero force, and the distances
$x_F$ and $x_U$ along the reaction coordinate that separates the transition state from states $F$ and $U$
respectively; the total distance along the reaction coordinate being
$x_m=x_F+x_U$. Under an applied force the free energy landscape is tilted along the reaction coordinate and the
free--energy difference $\Delta G$ and the barrier $B_0$ change accordingly. To a first approximation, $\Delta G$ and B depend linearly on the
force whereas $x_F$ and $x_U$ are taken force--independent. Hence the reaction rates are given by
\begin{eqnarray}
k_{F\rightarrow U}=k_0e^{-\beta \left(B-G_F-x_{FU}f\right)}=k_me^{\beta x_{\text{FU}f}}\label{capo}\\
k_{U\rightarrow F}=k_0e^{-\beta \left(B-G_U+x_{UF}f\right)}=k_me^{\beta\left(\Delta G_0-x_{\text{UF}}f\right)}\label{ope},
\end{eqnarray}
and $k_m=k_0e^{-\beta B_0}$ is an effective attempt rate.
The free--energy difference between state U and F under the given force $f$ is given by,
\begin{equation}\label{thermo}
 \Delta G(f)=-k_BT \log\left(\frac{k_{F\rightarrow U}}{k_{U\rightarrow F}}\right)=\Delta G_0-x_m f.
\end{equation}
with $\Delta G_{FU}(f)=G_F(f)-G_U(f)$.
The four parameters describing the free--energy landscape can therefore be reconstructed from the kinetic rates measured at different forces.

\end{document}